# Real-time Control System Prototype for Mechanical and Optical Systems based on Parallel Computing Techniques

F. Acernese[1,2], F. Barone[2,3], R. De Rosa[1,2], R. Esposito[2], P. Mastroserio[2], L. Milano[1,2], S. Pardi[1], K. Qipiani[2], F. Silvestri[1], G. Spadaccini[1,2]

[1]Università di Napoli "Federico II", Dipartimento di Scienze Fisiche, 80126 Napoli, Italy

[2]INFN – Sezione di Napoli, 80126 Napoli, Italy

[3]Università di Salerno, Dipartimento di Scienze Farmaceutiche, 84084 Salerno, Italy

Real-time control systems often require dedicated hardware and software, including real-time operating systems, while many systems are available for off-line computing, mainly based on standard system units (PCs), standard network connections (Ethernet), standard operating systems (Linux) and software independent from the particular architecture of the single unit. In order to try to get the advantages of both the technologies, we built an hybrid control system prototype using network based parallel computing architecture within real-time control system. In this paper we describe the architecture of the implemented system, the preliminary tests we performed for its characterization and the architecture of the control system we used for the real-time control tests.

## 1. INTRODUCTION

The design of a digital control system for a specific application requires the definition of the three main quantities: the control band, the signal dynamics, the computing power.

The control band is related both to the response speed of the control system and to the digital noise introduced in the data according to the sampling frequency [1]. Such requirements are often not too difficult to be satisfied for many classes of control problems. In particular, in the control of mechanical systems, a large variety of possible standards and low cost technologies, both hardware and software, are available. Their final choice depends generally on the definition of the other two parameters, that are the signal dynamics and the computing power. As an example, a digital system for the control of slow mechanical systems (10 Hz control band) can be not at all expensive if its architecture is based on a standard commercial Pc system with a free operating system like Linux.

On the other hand, in the case of digital control systems for the mechanical control of superattenuators for the locking of interferometric detectors of gravitational waves (10 Hz control band), the sampling rate must be quite large (10 kHz) in order to satisfy the requirement on the sensitivity of the system [2]. In the latter case, although still standard architectures are available, they are no more cheap and it is necessary to use real-time operating systems like VxWorks and LynxOS.

The acquisition section is also problem since the number of bits of the ADC and DAC is relevant both for the definition of the dynamics and of the sensitivity of the control system [1]. Very often the dynamic range of the acquisition, and, therefore, the number of the ADC and DAC bits, largely limits the acquisition frequency and hence the control band.

Finally, the computing power may be a very demanding requirement. In fact, although the number of input and output channel may be quite small, the control algorithm, like in the case of adaptive control algorithms, neural network based algorithms, optimization algorithms. A clear example is provided by adaptive optics control systems, whose control bands are of the order of 1 kHz, but with a number of input and output control channels quite large. The control of this system, although from a technical point of view well within the ADC and DAC technological limits, often requires very large computing powers [3].

As it is well known the implementation of a digital control system requires the definition of the requirements, the design of the architecture of the system and the choice of the hardware and the software. If the components are not available on the market, then it is necessary to develop ad hoc ones, generally not standard, not modular, often very expensive and difficult to maintain.

According to what said above, it may be of interest to try to find alternative solutions satisfying the requirements, but still standard, supported by the industry also in the developments and test phase, and, possibly, modular.

In the following section we will show how we implemented a digital control system using hybrid configurations, using the available technology and demonstrating the validity of an hybrid approach to the control of a suspended mass, part of the suspended Michelson interferometer developed in Napoli for the R&D on digital control systems for the development of gravitational waves interferometric detectors [4].

## 2. CONTROL SYSTEM ARCHITECTURE

The architecture of the digital control system of the suspended Michelson Interferometer is based on the VME standard (hardware). The operating system of all the VME CPUs is LynxOS, while all the software is written in standard C language. Within this digital control system architecture, we organized all the experimental tests on the hybrid digital control system architecture for the control of a suspension of one of the end mirrors.

### 2.1. The Mechanical System

As it is possible to see from Fig.1, the end mirror of the Interferometer is suspended to a two-stage seismic attenuator which simulate, in principle, the dynamics of a system with a marionetta and a test mass of the last stage of an interferometer for gravitational wave detection.

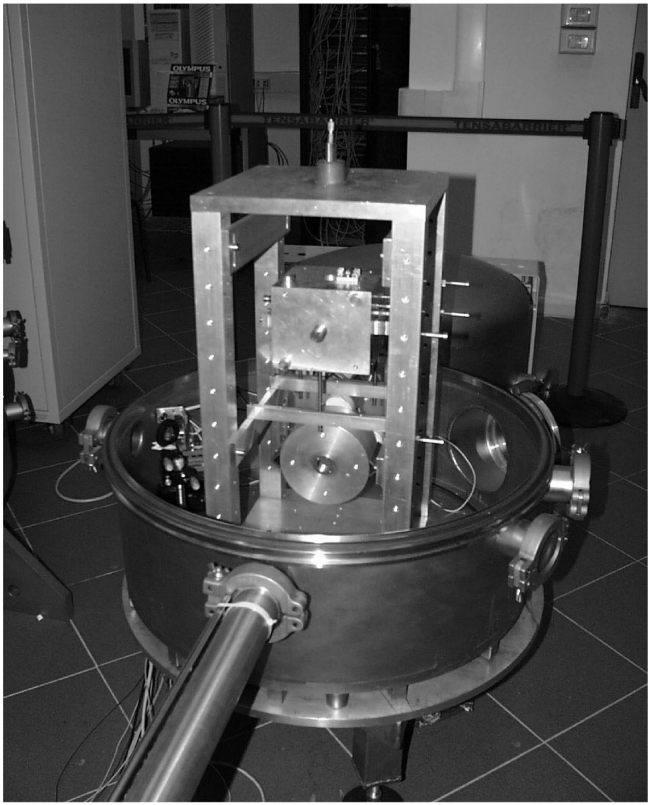

Fig.1 End Mirror Suspension

The first attenuation stage (*marionetta*) is an aluminium cube (14 cm side) barycentrically suspended with a steel wire of 0.2 mm diameter, while the mirror is mounted on the inner face of a cylindrical aluminium test mass (14 cm diameter, 6 cm heigth) suspended to the marionetta with four steel wires of 0.2 mm diameter.

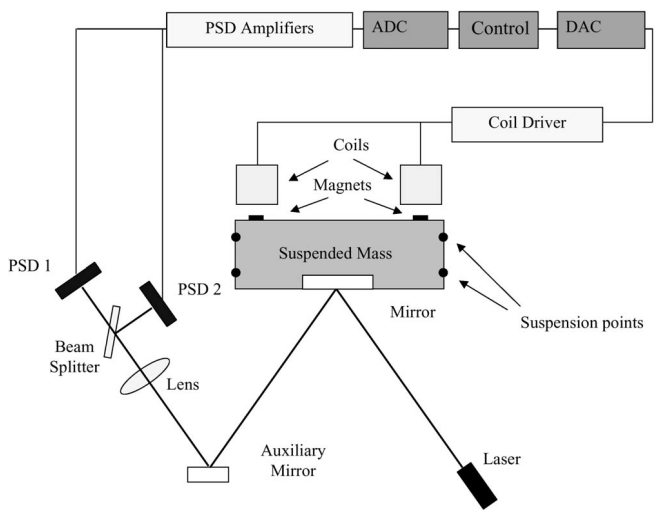

Fig.2 Control System of the last stage

Our control tests were performed on the control section devoted to the alignment control section of the suspension last stage (Fig.2).

## 2.2. The Digital System

The control system of the original interferometer digital control for the end mirror suspension consists of a VME CPU (VMPC4a from Cetia - CPU board with PowerPC 604 at 200 MHz, 64 MB RAM) running the operating system LynxOS and equipped with two 10 Mbit Ethernet links, interfaced to a 12bit ADC-DAC mod. VDAD from PEP Modular Computers.

Within this system we integrated our Computer Farm implemented for the development of off-line parallel data analysis of gravitational waves from coalescing binaries. The farm consists of 8 APPRO 2114Xi with Pentium IV 2.4 GHz and of 12 SuperMicro 6010H with Pentium III 1 GHz (see Fig.3). The operating system installed on the Napoli farm is Linux RedHat 7.3, kernel 2.4.20, with the OpenMosix extensions. This farms is also included within the GRID geographical network. In fact, each node of the farm is configured as a dual-boot system and can operate as a "grid-element" when needed.

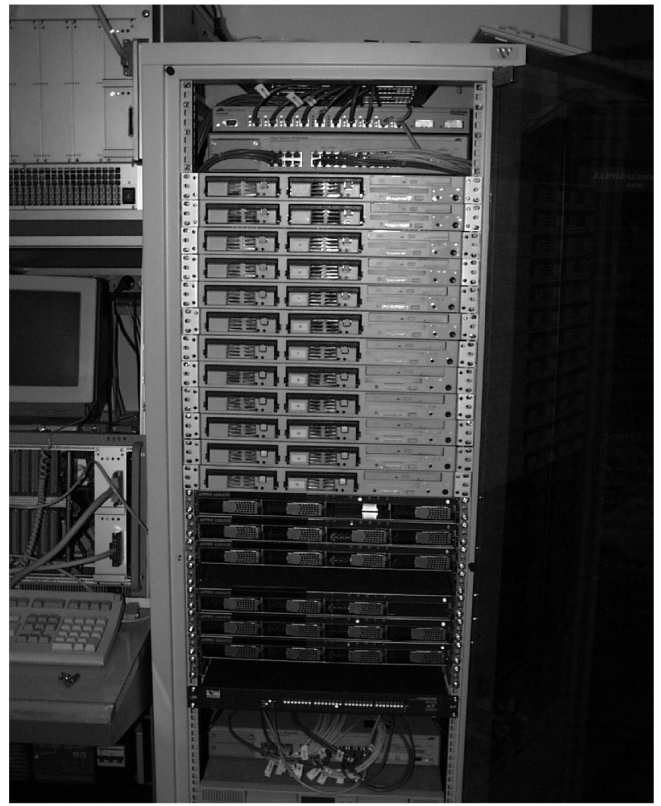

Fig.3 Napoli Computer Farm

Actually, in digital control systems the control is obtained by simply acquiring data at a sampling period (or equivalently at a sampling frequency) generally ten times the control band. Therefore, once that the sampling frequency is chosen, the control band of the system is

**TUGP003**

defined together with the maximum delay with which the acquired data (from ADC) must be presented at the output (DAC) after being processed for the generation of the control signal.

As a consequence, the most stringent requirement is to keep very stable and synchronous the sampling frequency (ADC frequency) and the conversion frequency (DAC frequency). The only requirement concerning the computation of the correction signal (control signal) is that it lasts always less than sampling period.

Therefore, from the control theory point of view, every asynchronous system may be considered synchronous if its response time, although changing in time, is always less than the sampling time. In our case this time is the sum of twice the data transfer time, $T_{DT}$, and of the computing time, $T_c$.

The hybrid architecture we have designed for the automatic control of the suspended interferometer is shown in Fig.4. The link between the farm (Linux) and real-time (LynxOS) architecture is made through the Ethernet network.

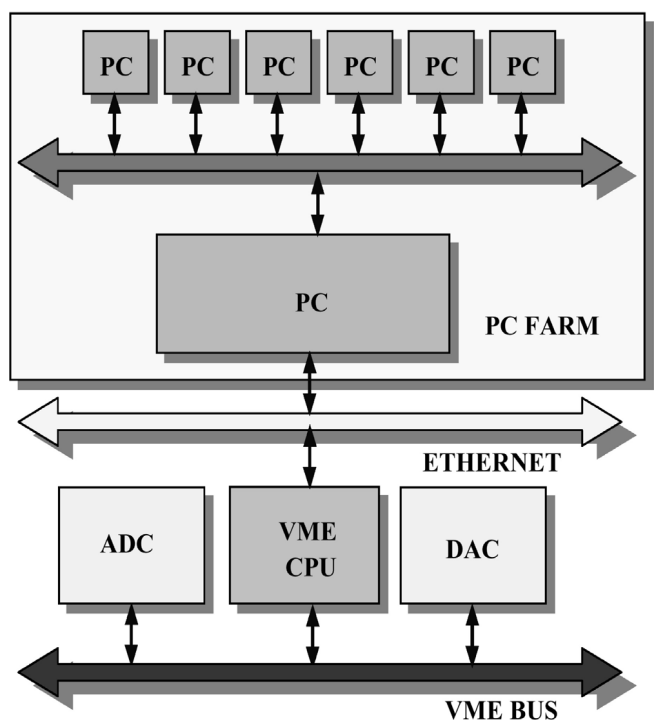

Fig.4 Hybrid Architecture for the Automatic Control of the Suspended Interferometer

## 3. TEST

In order to check the experimental feasibility of this architecture we organized some preliminary tests to evaluate the two quantities that put an upper limit to the control system band:

- data transfer time, $T_{DT}$
- computing power time, $T_C$.

**TUGP003**

### 3.1. Data Transfer Time

The first test we performed was the most important one, that is the data transfer between two PCs connected through a point-to-point Ethernet link.

In our case the test was performed using three different test protocols: the TCP/IP, the GAMMA (Genoa Active Message Machine) and the M-VIA (Virtual Interface Architecture). The results are shown in Table 1.

| Protocol | Max Time Delay |
|---|---|
| TCP/IP (eepro 100) | 55 us |
| TCP/IP (Gigabit) | 28 us |
| Gamma (eepro 100) | 33 us |
| M-VIA (DEC) | 23 us |

Table.1 Results of the Latency Tests

These results, although still provisional demonstrate that the data transfer time, $T_{DT}$, already allows the control system to sustain a sampling frequency of some tents of kHz. This value does not depend on the amplitude of the data transferred in our applications. In fact, assuming, for example, a digital control system with 100 inputs and output channels, each one sampled with 24 bit ADC, then the amplitude of each data block transferred with a single packet is only 300 byte, that has practically no influence on the global data transfer time.

Test on the full farm are still in progress, both on the possible parallel and serial computation configurations obtainable with its internal nodes.

### 3.2. Computing Power Time

The second test was related instead to the connection of the computing power available and the computing power time, $T_C$, that is the measurement of the time required for performing the computing operation for the generation of the control signal according to a control model to be sent to the actuators. This test was performed with one of the farm APPRO 2114Xi, whose performance was evaluated to be about 4 Gflops, according to the Pallas Benchmark. This result simply means, that in order to sustain a 10 kHz sampling rate, then the maximum rate of floating point operations must be less than $10^5$ Flops. This number of operations is well within the number of operations required by all the classical control algorithms for this class of problems and therefore the control system appears to be perfectly feasible.

## 4. MECHANICAL CONTROL RESULTS

In order to check the experimental feasibility of this approach we applied the basic architecture previously described to the alignment control of a suspended mass, replacing the existing classical real-time digital control.

The connection of an APPRO 2114Xi of the farm trough a dedicated point-to-point connection allowed us to sustain a control sampling frequency of 2.5 kHz for the above described digital control system actual configuration. The main difference between the original and the hybrid control system is that now it is available a large computing power and, therefore, more refined control algorithms can be easily implemented and tested.

The limitation in the sampling frequency to only 2.5 kHz is due to the data transfer time, that can be largely increased once that the VME CPU boards will be provided with a GigaEthernet connection. This upgrade will be tested in the next months.

## 5. CONCLUSIONS

We experimentally demonstrated that using standard hardware and software we were able to implement a control system for the alignment of a suspended mass with a sampling frequency of 2.5 kHz, using a computing power provided by a Pc master node of the farm external to the real-time control system.

These tests are very promising and encouraging in view of the extension of the computing power with the full inclusion of the internal nodes of the farm and to the upgrade of the VME CPU boards with GigaEthernet links.

Finally, it is very important to underline that the integration of such different techniques will make the control system more versatile and modular.